\documentclass[onecolumn,aps,prd,preprintnumbers,showpacs,superscriptaddress,nofootinbib,amsmath,amssymb,floats,floatfix,showkeys,notitlepage]{revtex4}

\usepackage{lipsum}
\usepackage{graphicx}
\usepackage{subfigure}
\usepackage{palatino}
\usepackage[utf8]{inputenc}
\usepackage{changes}
\usepackage{amssymb,amsmath,epsfig}
\usepackage{graphics}
\usepackage{hyperref}
\hypersetup{colorlinks=true,linkcolor=blue,urlcolor=blue,citecolor=blue}
\usepackage[toc,page]{appendix}
\usepackage[normalem]{ulem}
\usepackage{adjustbox}
\usepackage{latexsym}
\usepackage{amsmath}
\usepackage{amssymb}
\usepackage{amsfonts}
\usepackage{times}
\usepackage{dcolumn}
\usepackage{bm}
\usepackage{tikz}
\usepackage{bigints}
\usepackage{array,tabularx,multirow}
\usepackage[tracking=true]{microtype}
\usepackage{setspace}
\usepackage{listings}
\SetTracking{}{500}
\SetTracking{encoding={*}, shape=sc}{40}
\UseRawInputEncoding 
\allowdisplaybreaks
\begin{document} 

\title{Weak deflection angle by Kalb-Ramond Traversable Wormhole in Plasma and Dark Matter Mediums}

\author{Wajiha Javed}
\email{wajiha.javed@ue.edu.pk} 
\affiliation{Department of Mathematics, Division of Science and Technology, University of Education, Lahore-54590, Pakistan}

\author{Hafsa Irshad}
\email{hafsairshad989@gmail.com} 
\affiliation{Department of Mathematics, Division of Science and Technology, University of Education, Lahore-54590, Pakistan}

\author{Reggie C. Pantig}
\email{reggie.pantig@dlsu.edu.ph}
\affiliation{Physics Department, De La Salle University, 2401 Taft Avenue, Manila, 1004 Philippines}
\affiliation{Physics Department, Map\'ua University, 658 Muralla St., Intramuros, Manila 1002, Philippines}

\author{Ali {\"O}vg{\"u}n}
\email{ali.ovgun@emu.edu.tr}
\affiliation{Physics Department, Eastern Mediterranean University, Famagusta, 99628 North
Cyprus via Mersin 10, Turkey.}

\date{\today}

\begin{abstract}

 This paper is devoted to compute the weak deflection angle for Kalb-Ramond traversable wormhole solution in plasma and dark matter mediums by using the method of Gibbons and Werner. To acquire our results, we evaluate Gaussian optical curvature by utilizing the Gauss-Bonnet theorem in the weak field limits. We also investigate the graphical influence of deflection angle $\tilde{\alpha}$ with respect to the impact parameter $\sigma$ and minimal radius $r_0$ in plasma medium. Moreover, we derive the deflection angle by using different method known as Keeton and Petters method. We also examine that if we remove the effects of plasma and dark matter, the results become identical to that of non-plasma case. 
\end{abstract}


\date{\today}
\maketitle

\section{Introduction}

  A visual illustration of a black hole (BH) shows that it dissipates energy through radiation, then compresses and sooner or later dissolves \cite{Javed:2017saf}. Albert Einstein first estimated the presence of a BH by his theory of General Relativity \cite{Einstein}. A BH is considered as an area of space having too strong gravitational pull that the fastest moving objects even light cannot get out of it. Four types of BHs that exist are intermediate BHs, stellar BHs, miniature BHs and supermassive BHs. Black holes are certainly very simple as they have two primary parts, event horizon and the singularity. Event horizon is the boundary that indicates the limit of a BH. On event horizon, the escape velocity becomes equal to the velocity of light. While singularity is a point in space where the existing mass have infinite density. According to GR, spacetime singularities rise various issues, both scientifically and physically \cite{Javed:2020fli}.

Just like BHs, wormholes (WHs) appear as effective solutions to the Einstein field equations. The simplest solution to Einstein field equations is Schwarzschild solution. The idea of a WH was first given by Flamm in $1916$, soon after the invention of Schwarzschild's BH solution. In general, WH is the link between two separate regions of space that are far away from each other via a tunnel \cite{1}. In $1935$, Einstein and Rosen \cite{Einstein} additionally investigated the theory of inter-universe connections. These spacetime connections were known to be ``Einstein-Rosen Bridges". However, the idea of ``WH" was formulated by Wheeler in $1957$ \cite{Wheeler:1955zz,Fuller:1962zza}. After that, he illustrated that WH would be unsteady and non-traversable for even a photon.

Later on, the term traversable WH was developed by Morris and Thorne in 1988 \cite{Morris:1988cz}. They established flat traversable WHs with exotic matter that do not satisfy the null energy conditions \cite{Hochberg:1998ii}. Exotic matter creates problems for making stable WHs. Morris, Thorne and Yurtsever also showed that traversable WH can be made stable and flat by applying the Casimir effect. Another type of traversable WH, a thin-shell WH was introduced by Matt-Visser in 1989 \cite{Visser:1989kh} in which a path through the WH may be made in such a way that the traversing path does not cross the region of exotic matter. Also, the metric of Ellis WH was considered firstly in \cite{Ellis:1973yv} and analysis of Ellis WH returned to standard work of Morris and thorne where they introduced the traversable WHs. The deflection of light initially suggested by Chetouani and Clement in Ellis WH \cite{Chetouani}. Nakajima and Asada also studied Gravitational lensing by Ellis WH \cite{Nakajima:2012pu}.

 It is a known fact that weak and strong gravitational lensing (GL) is a very productive area to find not only dark and heavy objects but also BHs and WHs. To identify a WH, a possible way is the implementation of GL. The GL by WHs was reviewed on a large scale in the literature of theoretical physics as well as astrophysics \cite{Kuhfittig:2013hva}-\cite{31}.

Gravitational lensing suggested by Soldner for the first time in the background of Newtonian theory. The basic theory of GL is formed by Liebes, Refsdal and many other scientists \cite{32}-\cite{Pantig:2022gih}. When light released by distant galaxies, passed through the heavy objects in the universe, the gravitational attraction by those heavy objects can deviate the light from its pathway. This process is known as GL. Gravitational lensing is very applicable method to understand dark matter, galaxies and universe. Three types of GL listed in the literature are: (i) strong GL (ii) weak GL (iii) micro-GL. The strong GL indicates the approximate magnification, location and time delay of the images with the aid of BHs. Strong GL is also helpful to see various objects such as boson stars, fermion stars and monopoles \cite{Barriola}. The weak GL is used to discover the mass of astronomical objects without demanding their changing nature. Weak lensing also differentiates dark energy from modified gravity and investigates how universe is expanding rapidly. Gravitational lensing has been calculated for many spacetimes by applying various methods \cite{19}-\cite{Gibbons:2008rj}. In the past few years, we examine many researches that relate GL with the Gauss-Bonnet theorem (GBT).

Using GBT, Gibbons and Werner \cite{Gibbons:2008rj} revealed that it is possible to compute the deflection angle in weak field limits using Gauss-bonnet theorem. After that, Werner expanded this technique to Kerr BHs \cite{Werner:2012rc} by using Nazim's method for Rander-Finsler metric. Then, Ono et al. enlarged the study of weak GL in stationary axisymmetric spacetimes using finite distance method \cite{42}-\cite{Takizawa:2020egm}.
The theory of GL comprises of three physical processes named as (i) geometric optics (ii) thin lens approximation  (iii) perturbation theory of GL \cite{Keeton:2005jd,Keeton:2006sa,Sereno}.

For a distant observer from a source, the bending angle of light can be determined by using GBT in the weak field limits \cite{Gibbons:2008rj}. By considering an oriented surface, let us describe a domain $D_R$ surrounded by the beam of light with a circular boundary $C_R$ having Euler characteristic element $\mathcal{X}$  and metric $g$ at the focus area where light rays coincide with source and the viewer. So, when the GBT is applied within the optical metric, it provides us the bending angle of light stated as \cite{Gibbons:2008rj}:

\begin{equation}
 \int\int_{\mathcal{D}_{R}}\mathcal{K}dS+\oint_{\partial\mathcal{D}_{R}}kdt
 +\sum_{n}\theta_{n}=2\pi\mathcal{X}(\mathcal{D}_{R}),\nonumber\\
\end{equation}

where, $D_R$ is the region that comprises of the source of light waves, observer and the focal point of the lens, $\mathcal{K}$ shows the Gaussian optical curvature, $dS$ is surface element, $k$ is known as geodesic curvature and $\partial\mathcal{D}_{R}$ shows that this portion is surrounded by the outermost light rays. The asymptotic bending angle of light can be computed as \cite{Gibbons:2008rj}:

\begin{equation}
\tilde{\alpha}=-\int \int_{D_\infty} \mathcal{K} dS,\nonumber\\
\end{equation}

where $\tilde{\alpha}$ represents the bending angle and $D_\infty$ indicates the infinite domain bounded by the rays of photon, apart from the lens.
Moreover, By utilizing GBT, bending angle for static and axisymmetric rotating Teo WH was investigated by Jusufi and {\"O}vg{\"u}n \cite{JusufiOvgun}. Recently, \"Ovg\"un has studied deflection angle by Damour-Solodukhin WHs \cite{Ovgun:2018fnk}.

An approximate form of GL by the help of spherically symmetric lenses upto post-post-Newtonian (PPN) is newly originated by Keeton and Petters \cite{Keeton:2005jd,Keeton:2006sa}. Sereno and de Luca \cite{Sereno} expanded the PPN approximation to Kerr BHs. The Keeton and Petters method allows the calculations of those observable quantities that can be basically independent of coordinates and therefore physically applicable.

Dark matter comprises of those particles that do not absorb, reflect, or emit light, so that they cannot be detected by observing electromagnetic radiations. Dark matter is the substance that cannot visualize directly. Dark matter produces upto $27\%$ of overall mass-energy of the universe and the remaining part consists of dark energy. Dark matter can only be noticed by their gravitational interaction. Although, they have weak non-gravitational interaction and are of non-relativistic nature \cite{Feng}. Some of the dark matters exist are: weakly relating massive particles (WIMPs), axions, sterile neutrinos, super WIMPs etc. Dark matter particles are formed by non-baryonic particles. The neutrino was the only familiar non-baryonic particle and it is taken as the first dark matter candidate. Refractive index used in dark matter maintains the propagation speed.
According to the $20^{th}$ century report, we have seen a lot of times that mostly matter consists of protons and neutrons. But that matter which we see is not suitable form of matter. In this universe, there are some other types of matter whose masses are five times greater than the regular matter. Such unknown matter is named as ``dark matter". We cannot detect its existence in the laboratory till now. For the proper understanding of dark matter, we have to understand and then use various branches of astronomy and physics such as particle physics is important in describing the relation between dark matter and standard matter. Cosmology, General Relativity and astrophysics are applicable for the study of dark matter on a wide range in literature \cite{Latimer:2013rja,Latimer:2017lwm}.
For the inspection of bending angle through dark matter, we take the refractive index as \cite{Latimer:2013rja,Latimer:2017lwm}:
\begin{equation}
n(\omega)= 1+ {\beta} A_0+ A_2 {\omega}^2,\nonumber\\
\end{equation}
where, $\omega$ is the frequency of photon. It is noticed here that $\beta= \frac{\rho_0}{4m^2 \omega^2}$ represents the mass density of the scattered dark matter particles, $A_0=-2e^2\epsilon^2$ and $A_2j\geq 0$. The higher order terms like $O(\omega^2)$ and onward are linked with polarizability of the dark matter candidate. The terms $O(\omega^2)$ stand for the neutral dark matter candidate while $O(\omega^{-2})$ denote the charged dark matter candidates.

The main goal of this paper is to examine the deflection of light for Kalb-Ramond traversable WH solution by using various methods.

This paper is organized as; In section $\textbf{2}$, we analyze the Kalb-Ramond traversable WH solution in detail. In section $\textbf{3}$, we compute the deflection angle for Kalb-Ramond traversable WH solution in plasma medium. Section $\textbf{4}$, is related to the graphical inspection of deflection angle of Kalb-Ramond traversable WH solution in the framework of plasma medium. In section $\textbf{5}$, we investigate the bending angle of Kalb-Ramond traversable WH solution by using Keeton and Petters method. In section $\textbf{6}$, we enlarge our observations and derive the value of deflection angle in dark matter medium. In section $\textbf{7}$, we conclude all the results.

\section{Kalb-Ramond Traversable Wormhole Solution}
This section is devoted to describe some properties of a traversable WH solution. The metric is spherically symmetric and independent of time for the sake of simplification. Wormhole is like a tunnel connecting two regions of space, have no horizon because the presence of horizon does not allow two-sided travelling. The time taken by the traveller to pass through WH should be limited and fairly short. The traveller experiencing the gravitational force inside the WH should be slightly small.\\
Local Lorentz violation effects also result in modified gravitational dynamics. Although the Lorentz symmetry may be violated close to the Planck scale. When one or more of the tensor fields gain nonzero vacuum expectation values (VEV), Lorentz symmetry breaks spontaneously. The Kalb-Ramond (KR) field, an antisymmetric tensor field $B\mu\nu$ that appears in string theories, is another cause for spontaneous Lorentz symmetry breaking (LSB). A non-vanishing VEV breaks the gauge and the Lorentz symmetry by allowing a self-interaction potential. The tensor antisymmetric VEV can be split into two vectors, called pseudo-electric and pseudo-magnetic vectors, in the same way that the electromagnetic field strength can. A background KR field recently altered a static and spherically symmetric wormhole. The parameter-dependent power-law correction to the Schwarszchild solution resulted from a non-minimal interaction between the KR VEV and the Ricci tensor. The effects of this LSB solution on the WH temperature and the shadows were also investigated \cite{Lessa}.
The static spherically symmetric spacetime of a Morris-Thorne traversable WH is expressed by the metric \cite{54}:
\begin{equation}
ds^2=-e^{2{\phi(r)}}dt^2 + \frac{dr^2}{1-\frac{\Omega(r)}{r}} + r^2 d\theta^2 + r^2\sin^2\theta d\phi^2,\label{HF1}
\end{equation}
where the value of $\Omega(r)$ is defined as: \cite{Lessa:2020imi}
\begin{equation}
\Omega(r)=r(\frac{r}{r_0})^\frac{2}{1-2\lambda}.\nonumber
\end{equation}
Here, $\lambda$ is the Lorentz violating parameter $\lambda=
|b|^{2} \xi_{2}
$ , $b$ and $\xi_{2}$ are constants \cite{Lessa:2020imi}, $\phi(r)$ is redshift function and $\Omega(r)$ is the shape function of WH. Both of these functions are adjustable. Redshift function is finite everywhere. The shape function $\Omega(r)$ calculates the shape of the WH or describes the WH physically. 
The radial variable $r$ has a minimum value of $r_0$ and a maximum value of infinity, or $r_0 \leq r< \infty$. The $r_0$ is a positive constant as it 
decreases from positive infinity to a least value and after that starts moving towards positive infinity.

After putting the value of shape function, the spacetime metric  becomes:
\begin{equation}
ds^2=-dt^2 + \frac{dr^2}{1-(\frac{r}{r_0})^\frac{2}{1-2\lambda}} + r^2 (d\theta^2 + \sin^2\theta d\phi^2).\label{HF3}
\end{equation}
Now, for Kalb-Ramond traversable WH solution, we can write our metric in general form as:
\begin{equation}
ds^2=-g_{tt}dt^2+ g_{rr}dr^2+ r^2 d\theta^2+ r^2 \sin^2\theta d\phi^2, \label{HF4}
\end{equation}
where, $g_{tt}=1$ and $g_{rr}=\frac{1}{1-(\frac{r}{r_0})^\frac{2}{1-2\lambda}}$. For the calculation of deflection angle by Kalb-Ramond traversable WH solution, the optical path can be obtained by the null geodesic condition $ds^2 = 0$. Also, we assume that the photon is moveable in equatorial plane ($\theta=\frac{\pi}{2}$). In order to calculate the deflection angle, one can obtain the optical path metric in explicit form given as follows: \cite{Lessa:2020imi}
\begin{equation}
dt^2=\frac{dr^2}{1-(\frac{r}{r_0})^\frac{2}{1-2\lambda}} + r^2 d\phi^2. \label{HF5}
\end{equation}
The above metric will be used to calculate the deflection angle of light by Kalb-Ramond travesable WH solution which is described.
\section{Deflection Angle in Plasma Medium}

In this section, we evaluate impact of plasma medium on the weak GL for Kalb-Ramond traversable WH solution (\ref{HF5}). The value of refractive index $n(r)$ for the given WH solution is defined as \cite{55}:
\begin{equation}
n(r)=\sqrt{1-\frac{\omega_e^2(r)}{\omega_\infty^2(r)}f(r)}. \label{HF6}
\end{equation}
In refractive index $n(r)$, $\omega_e$ indicates the plasma frequency of electron while $\omega_\infty$ denotes the frequency of photon that is noticed by the observer at infinity. So, optical metric can be written as:
 \begin{equation}
 dt^2=g^{opt}_{lm}dx^ldx^m = n^2(r)\left[{\frac{dr^2}{f(r)}+r^2d\phi^2}\right],\label{HF7}
\end{equation}
where, $f(r)$ is defined as
\begin{equation}
f(r)=\frac{1}{1-(\frac{r}{r_0})^\frac{2}{1-2\lambda}}.\label{HF10}
\end{equation}
Now, the relative Guassian optical curvature can be determined by the expression present in \cite{57,58}:
\begin{equation}
\mathcal{K}=\frac{R}{2},\label{HF8}
\end{equation}
where $R$ is Ricci scalar. We calculate the Guassian optical curvature for Kalb Ramond traversable WH solution as:
\begin{equation}
\mathcal{K}\simeq-\frac{r_0}{2r^3}+\frac{5}{4}\frac{r_0^2}{r^4}\frac{w_{e}^2}{w_{\infty}^2}
-\frac{r_0}{r^3}\frac{w_{e}^2}{w_{\infty}^2}+\mathcal{O}({r_{0}^3}).\label{HF8}
\end{equation}
The Guassian curvature depends on minimal radius $r_0$ and radial parameter $r$. For the sake of simplicity, we consider the optical curvature upto order $2$ as well as to match our results given in non-plasma medium case.

The infinitesimal surface element for Kalb-Ramond traversable WH solution can be computed as:
\begin{equation}
dS=\sqrt {g}  dr d\phi= \left(r-\frac{r w_{e}^2}{w_{\infty}^2}\right)drd\phi+\mathcal{O}({r_{0}^3}).\nonumber
\end{equation}
To find the deflection angle in plasma medium, we utilize the GBT. As the beam of light approaches from the infinity upto a large distance and remember that we are present in weak field limitations. Here, beam of light becomes almost straight. Hence, we utilize the straight line approximation  $r= \frac{\sigma}{\sin\phi}$, where $\sigma$ expresses the impact parameter, GBT is stated as \cite{Gibbons:2008rj}:
\begin{equation}
\tilde{\alpha}=-\int_{0} ^{\pi} \int_\frac{\sigma}{\sin\phi} ^{\infty} \mathcal{K}dS.\label{HF13}
 \end{equation}

The resulting expression can be expressed for $\lambda= \frac{3}{2}$. The Lorentz violating parameter must be restricted to the range $\lambda > 1/2$ to achieve an asymptotically flat spacetime. If we take $\lambda\simeq 0$, the geometry of metric will not be asymptotically flat.
For this value of $\lambda$, deflection angle $\tilde{\alpha}$ for Kalb-Ramond traversable WH can be obtained as:
\begin{equation}
\tilde{\alpha}\simeq \frac{r_0}{\sigma} + \pi \left(\frac{r_0}{4\sigma}\right)^2+ \frac{w_{e}^2}{w_{\infty}^2} \frac{r_0}{\sigma} + \frac{r_0^2 \pi}{8\sigma^2} \frac{w_{e}^2}{w_{\infty}^2}+\mathcal{O}({r_{0}^3}).\label{MK}
 \end{equation}
The above result \label{HF16} shows that the light rays are moving in plasma medium. If we remove plasma effects, this angle will convert into the non-plasma medium. The deflection angle $\tilde{\alpha}$ depends on minimal radius $r_0$ and the impact parameter $\sigma$, where bending angle is directly proportional to $r_0$ and inversely proportional to $\sigma$.

\section{Graphical Behavior of Deflection Angle}

This section is based on the explanation of the graphical analysis of deflection angle for Kalb-Ramond traversable WH solution in plasma medium. For this purpose, we analyze the deflection angle with respect to the impact parameter $\sigma$ and the minimal radius $r_0$. 

\subsection {$\tilde{\alpha}$ versus $\sigma$}

\begin{center}
\epsfig{file=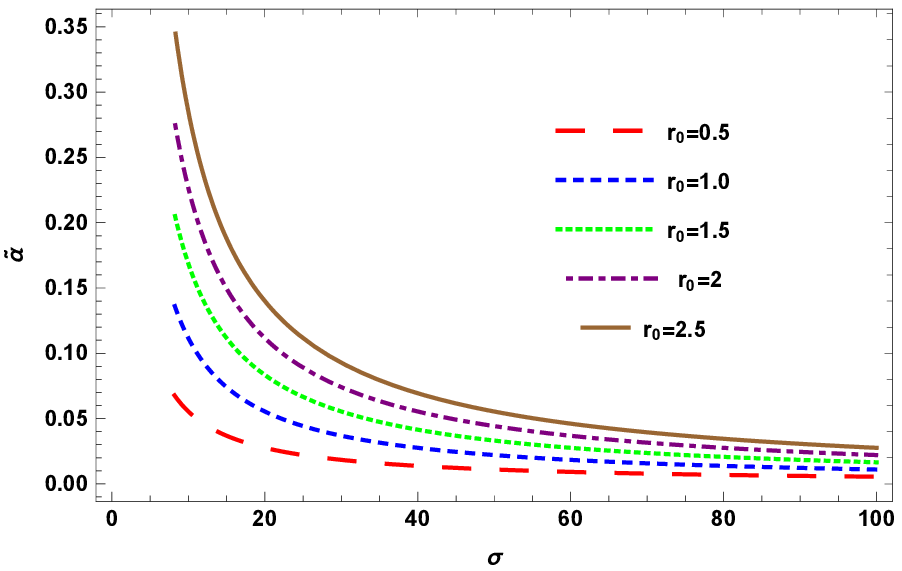,width=0.50\linewidth}\epsfig{file=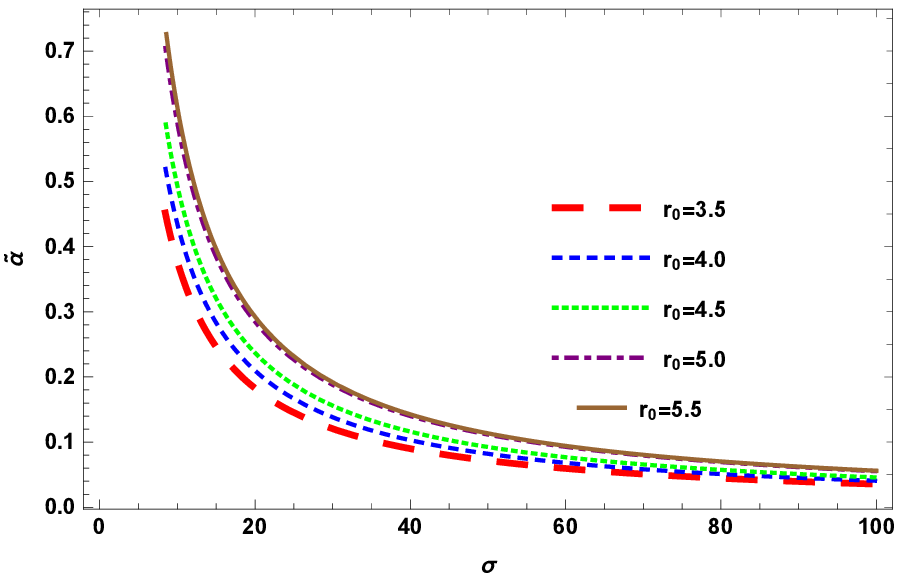,width=0.50\linewidth}\\
{Figure 1: $\tilde{\alpha}$ versus $\sigma$ }.
\end{center}
{Figure 1} shows the behavior of $\tilde{\alpha}$ with respect to the impact parameter $\sigma$ for plasma impact $\frac{\omega_e}{\omega_\infty}=10^{-1}$. For $r_0\leq3$, we observe that, in first graph as $\sigma$ increases, $\tilde{\alpha}$ decreases exponentially and shows a convergent behavior; which converges to zero. While on the other hand, we evaluate that with the increase of $r_0$, the deflection angle also increases. Similarly for $r_0\geq3$ in second plot, the angle shows a decreasing behavior when value of impact parameter increases.
\begin{center}
\epsfig{file=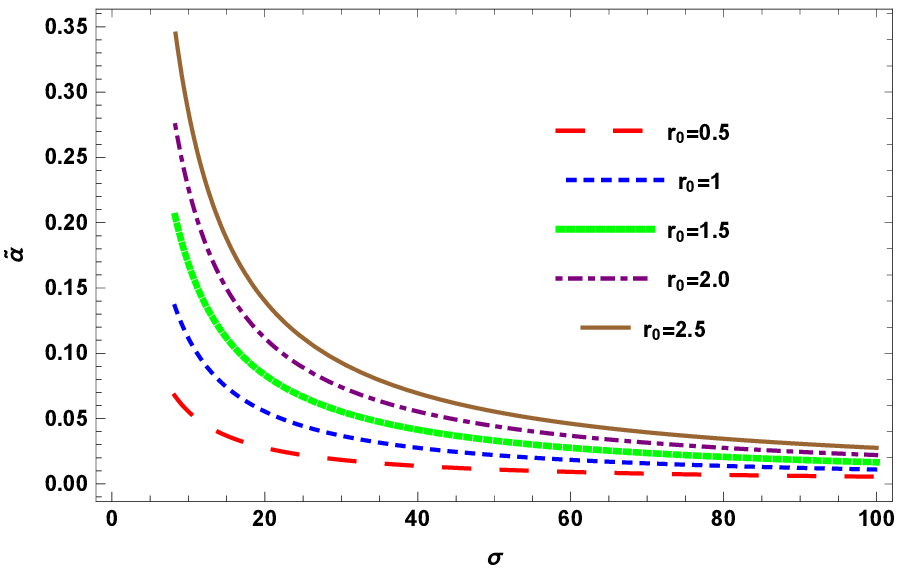,width=0.50\linewidth}\epsfig{file=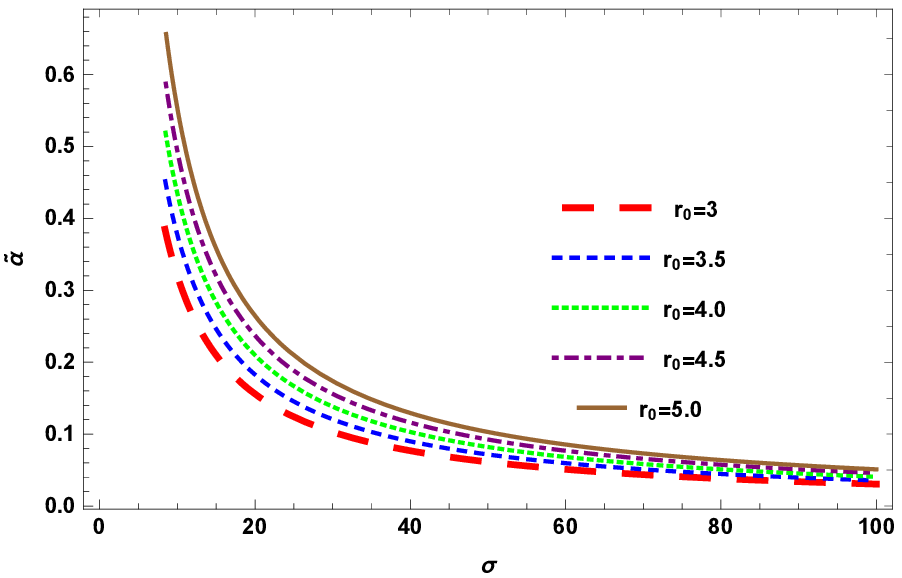,width=0.50\linewidth}\\
{Figure 2: $\tilde{\alpha}$ versus $\sigma$}.
\end{center}
Figure 2 shows that for $r_0\leq3$, $r_0>3$ by assuming plasma impact $\frac{\omega_e}{\omega_\infty}=10^{-2}$, one can analyze the similar behavior like that of $\frac{\omega_e}{\omega_\infty}=10^{-1}$. We also investigate that for small values of $r_0$, as $\sigma$ increases, $\tilde{\alpha}$ exponentially decreases. Moreover, it is to be observed that as the value of $\sigma$ increases, $\tilde{\alpha}$ decreases. It shows that deflection angle has inverse relation with $\sigma$ and direct relation with $r_0$.

\subsection{ $\tilde{\alpha}$ versus $r_0$}
\begin{center}
\epsfig{file=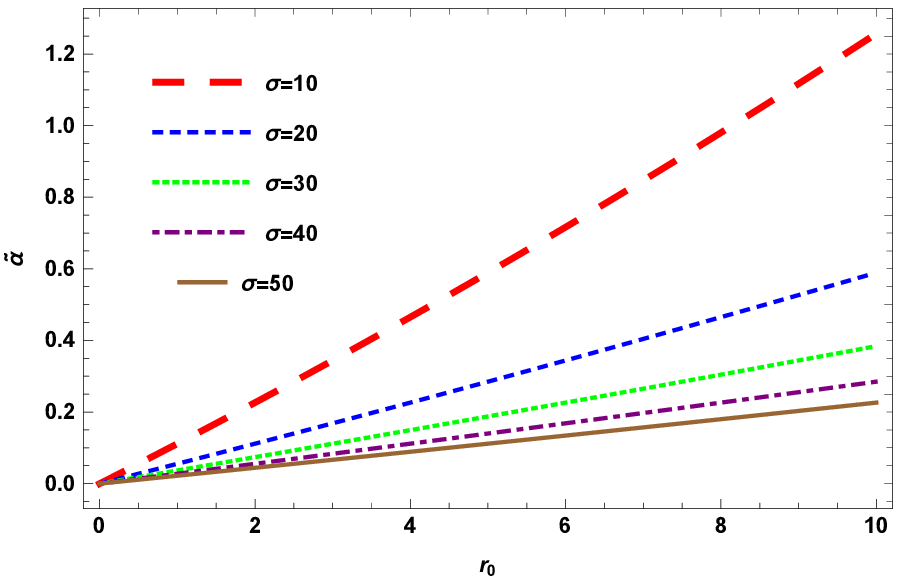,width=0.50\linewidth}\epsfig{file=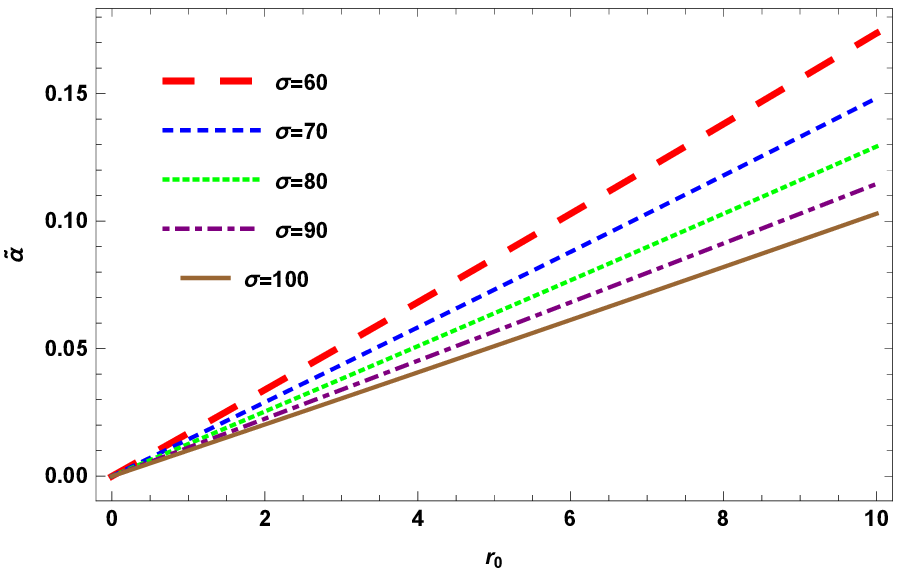,width=0.50\linewidth}\\
{Figure 3: $\tilde{\alpha}$ versus $r_0$}.
\end{center}

Figure 3 shows the behavior of deflection angle $\tilde{\alpha}$ of light with respect to $r_0$ for the impact parameter $0<\sigma< 50$ and plasma impact $\frac{\omega_e}{\omega_\infty}=10^{-1}$. We examine that for $0<\sigma<50$, the left graph indicates the linear behavior as $r_0$ increases the bending angle $\tilde{\alpha}$ also increases and as $\sigma$ increases, deflection angle decreases. For larger $\sigma$ angle approaches towards zero. The similar case is with the right plot for $\sigma>50$ that angle increases with increasing the values of $r_0$ and vice versa.
\begin{center}
\epsfig{file=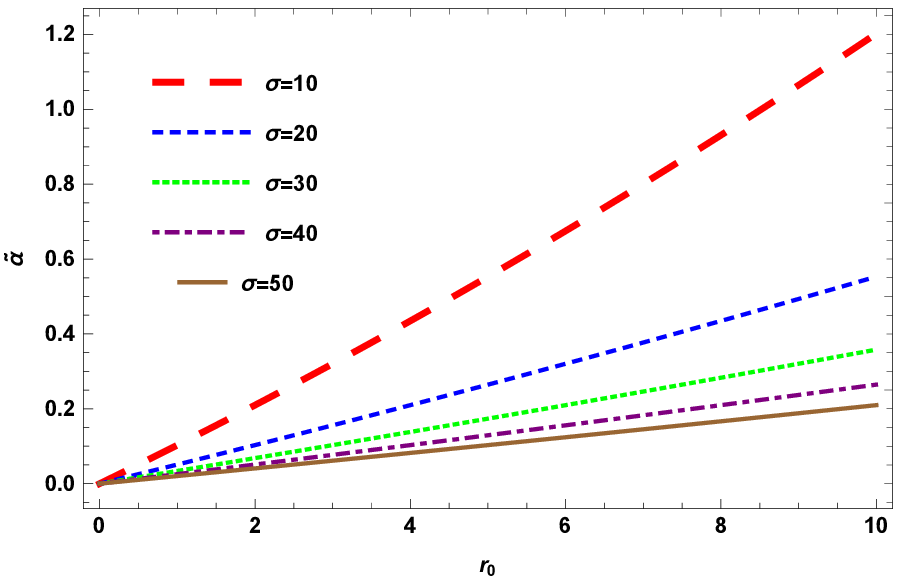,width=0.50\linewidth}\epsfig{file=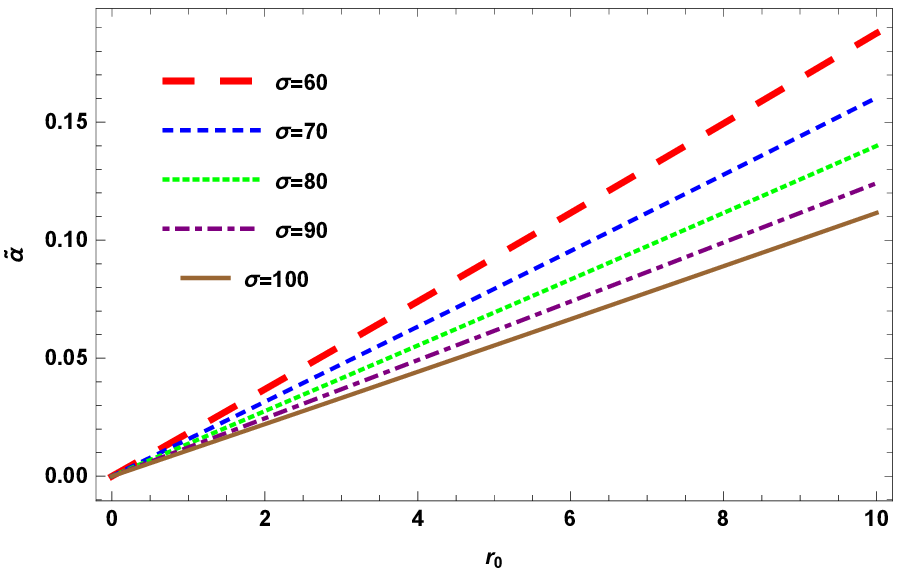,width=0.50\linewidth}\\
{Figure 4: $\tilde{\alpha}$ versus $r_0$}.
\end{center}

Figure 4 shows that for $\sigma>50$ and $0<\sigma\leq50$ having plasma impact's value $\frac{\omega_e}{\omega_\infty}=10^{-2}$, we can analyze similar behavior as for case $\frac{\omega_e}{\omega_\infty}=10^{-1}$. We evaluate that as values of $r_0$ increases, deflection angle also increases and as $\sigma$ increases, bending angle decreases which shows a divergent behavior of graphs. Furthermore, we investigate that bending angle $\tilde{\alpha}$ has direct relation with impact parameter $r_0$ and inverse relation with $\sigma$.

\section{Deflection Angle using Keeton and Petters Method}

 Keeton and Petters established a completely beneficial framework for computing corrections in a standard asymptotically flat metric theory of gravity \cite{Keeton:2005jd,Keeton:2006sa}. The central focus is to illustrate a way to manage lensing in computing gravity theories using PPN corrections upto third order.
 
The non-linear spherically symmetric and asymptotically Minkowski spacetime is defined by:
\begin{equation}
ds^2=-A\left(r\right)dt^2+B\left(r\right)dr^2+C\left(r\right)d\Omega^2.\label{HF17}
\end{equation}
Due to the spherical symmetry, geodesics of Eq.(\ref{HF17}) lie in equatorial plane while $d\Omega^2=d\theta^2 + \sin^2\theta d\phi^2$ represents the standard unit metric. The above equation becomes:
\begin{equation}
ds^2=-A\left(r\right)dt^2+B\left(r\right)dr^2+C\left(r\right)d\theta^2.
\end{equation}
For $A\left(r\right)\rightarrow 1$,  $B\left(r\right)\rightarrow 1$ and  $C\left(r\right)\rightarrow r^2$, the spacetime metric is flat in the absence of lens and we suppose that $A$, $B$ and $C$  are all +ve in the outside region of the lens from where the required light rays pass.
The Kalb-Ramond traversable WH solution can be presented as:
\begin{equation}
ds^2=-dt^2 + \frac{dr^2}{1-(\frac{r}{r_0})^\frac{2}{1-2\lambda}} + r^2 d\theta^2,\label{HF18}
\end{equation}
where the metric functions are:
\begin{equation}
A\left(r\right)= 1,~~~~~~\label{HF19}
\end{equation}
and
\begin{equation}
~~~~~~~\\B\left(r\right)= \frac{1}{1-(\frac{r}{r_0})^\frac{2}{1-2\lambda}}.\label{HF20}
\end{equation}
To find the PPN coefficients, we compare the coefficients of the extended form of metric function with the coefficients of standard form of general metric in a PPN series to third order. The general form of PPN series for Eq.(5.1) given as \cite{Keeton:2005jd}:
\begin{equation}
A(r)=1+2a_1\left(\frac{\phi}{c^2}\right)+2a_2\left(\frac{\phi}{c^2}\right)^2 +2a_3\left(\frac{\phi}{c^2}\right)^3 ...\label{HF23}
\end{equation}
\begin{equation}
B(r)=1-2b_1\left(\frac{\phi}{c^2}\right)+4b_2\left(\frac{\phi}{c^2}\right)^2 -8b_3\left(\frac{\phi}{c^2}\right)^3 ...\label{HF23}
\end{equation}
Here, $\phi$ denotes the three dimensional Newtonian prospective. In our paper, we consider,
\begin{equation}
\left(\frac{\phi}{c^2}\right)=\frac{r_0}{r}.
\end{equation}
We get the values of the coefficients of PPN metric as
\begin{equation}
a_1=0,~~a_2=0,~~a_3=0,~~ b_1=\frac{1}{2},~~b_2=\frac{1}{4},~~b_3=\frac{1}{8}.\nonumber
\end{equation}
After finding the PPN coefficients, we then determine the coefficients in the extended form of the bending angle. The extended form of light deflection angle can be written as follows
\begin{equation}
\alpha\left(b\right)=A_1\left(\frac{m}{b}\right)+A_2\left(\frac{m}{b}\right)^2 +A_3\left(\frac{m}{b}\right)^3 +O\left(\frac{m}{b}\right)^4.\label{HF23}
\end{equation}
We take $r_0=m$ and $\sigma=b$. To compute the coefficients of bending angle, we utilize the following equation
\begin{eqnarray}
A_{1}&=& 2\left(a_1+b_1\right),\nonumber
\\A_{2}&=&\left(2a_1^2\ -a_2+a_1 b_1+b_2-\frac{b_1^2}{4}\right)\pi,\nonumber
\\A_{3}&=&\frac{2}{3}[35a_1^3 +15a_1^2 b_1 -3a_1\left(10a_2 +b_1^2 -4b_2\right)\nonumber
\\ &+&6a_3 +b_1^3 -6a_2 b_1 -4b_1 b_2 +8b_3].\label{HF26}
\end{eqnarray}
Using the Eq.(\ref{HF26}), we can determine the values of coefficient presented as follows
\begin{equation}
A_1=1,~~~~~A_2=\frac{3\pi}{16}~~~~,A_3=\frac{5}{12}.\nonumber
\end{equation}
After substituting the values of coefficients in Eq.(\ref{HF23}), we get the resulting bending angle of Kalb-Ramond traversable WH as:
\begin{equation}
\tilde{\alpha}\left(\sigma\right)= \left(\frac{r_0}{\sigma}\right) +\frac{3\pi}{16} \left(\frac{r_0}{\sigma}\right)^2 +\frac{5}{12} \left(\frac{r_0}{\sigma}\right)^3+ O\left(\frac{r_0}{\sigma}\right)^4.\label{HF28}
\end{equation}

This expression shows the deflection angle by using Keeton and Petters method which is dependent upon $r_0$ and $\sigma$. The bending angle is directly proportional to $r_0$ and inversely proportional to $\sigma$.

\section{Deflection Angle of Photon in Dark Matter Medium}

Here, in this section, we examine how dark matter influences the weak deflection angle. For this purpose, we take the refractive index for dark matter medium \cite{Latimer:2013rja,Latimer:2017lwm,Ovgun:2020yuv}:
\begin{equation}
n(\omega)= 1+ {\beta} A_0+ A_2 {\omega}^2.\label{HF29}
\end{equation}

The 2-dimensional optical geometry for Kalb-Ramond traversable WH solution is:
\begin{equation}
dt^2=n^2\left(\frac{dr^2}{1-(\frac{r}{r_0})^\frac{2}{1-2\lambda}} + r^2 d\phi^2\right),\label{HF30}
\end{equation}
with the condition:
\begin{equation}
 \frac{dt}{d\phi}|_{C_R} = n\left(r^2 \right)^\frac{1}{2}.\label{HF31}
\end{equation}
Applying this condition to a non-singular domain $C_R$ outside of the light ray will yield the deflection angle.
Hence, with the help of GBT we can calculate the weak deflection angle for Kalb-Ramond traversable WH solution in dark matter medium.
\begin{equation}
\lim_{R\rightarrow \infty}\int_{0} ^{\pi+ \alpha} \left[\mathcal{K}_{g} \frac{dt}{d\phi}\right]|_{C_R} d\phi  =\pi-\lim_{R\rightarrow \infty}\int \int_{D_R}  \mathcal{K} dS.\label{HF32}
\end{equation}
We compute the Guassian optical curvature as follows:
\begin{equation}
\mathcal{K}= \frac{r_0}{2 r_3\left(1+ {\beta} A_0+ A_2 {\omega}^2\right)}.\label{HF33}
\end{equation}
After this, we determine
\begin{equation}
\lim_{R\rightarrow \infty} \mathcal{K}_{g} \frac{dt}{d\phi}|_{C_R}= 1.\label{HF34}
\end{equation}
Now, when we apply limit $R\rightarrow\infty$, then we can evaluate the deflection angle for Kalb-Ramond traversable WH solution by using GBT as below
\begin{equation}
\tilde{\alpha}=-\int_{0} ^{\pi} \int_\frac{\sigma}{\sin\phi} ^{\infty} \mathcal{K} dS.\label{HF35}
 \end{equation}
We get the weak deflection angle of Kalb-Ramond traversable WH solution in dark matter medium after putting the value of $\mathcal{K}$ and $dS$

\begin{equation}
\tilde{\alpha}\simeq \frac{r_0}{\sigma \left(1+ \beta A_0+ A_2 \omega^2\right)^6}+ \frac{r_0^2 \beta^6 A_0^{12}  \pi}{16\sigma^2 \left(1+ \beta A_0+ A_2 \omega^2 \right)^6}|+\mathcal{O}(r_{0}^{3}).\label{HF36}
\end{equation}

This result shows that when the effect of dark matter vanishes, the bending angle reduces to the deflection angle in non-plasma medium. The effect of dark matter shows large deflection than in general case. The deflection angle $\tilde{\alpha}$ depends upon $\sigma$ and $r_0$. We can see that dark matter gives a large deflection angle than in general case.
\section{Conclusion}

This paper is concerned to calculate the bending angle $\tilde{\alpha}$ for Kalb-Ramond traversable WH solution in plasma and dark matter mediums. We have evaluated the metric and explained its mathematical interpretation. For the calculation of bending angle, we have obtained the Guassian optical curvature. After that, we applied GBT and derived the value of deflection angle for Kalb-Ramond traversable WH solution in plasma medium Eq.(\ref{MK}) for $\lambda=\frac{3}{2}$.

If we neglect or approach the value of the plasma effect $\frac{\omega_e}{\omega_\infty}$ to zero, then effect of the plasma vanished and the resulting expression of bending angle for plasma medium reduced to the non-plasma case.

Graphical behavior of deflection angle in plasma medium has also been calculated.

\textit{\textbf{$\tilde{\alpha}$ with respect to impact parameter $\sigma$:}}

For $r_0\leq3$ and $r_0>3$ having effect of plasma $\frac{\omega_e}{\omega_\infty}=10^{-1}$, the deflection angle $\tilde{\alpha}$ gradually approaches towards zero as $\sigma$ increases, in both graphs. We have also observed that $\tilde{\alpha}$ exponentially decreases and indicate a convergent behavior for large values of $\sigma$.
 We have examined that for $r_0\leq3$ and $r_0>3$ having  plasma effect $\frac{\omega_e}{\omega_\infty}=10^{-2}$ as $r_0$ approaches to infinity, $\tilde{\alpha}\rightarrow\infty$ at smaller $\sigma$. We have also investigated that $\tilde{\alpha}$ has inverse relation with $\sigma$ and direct relation with $r_0$ .

\textit{\textbf{$\tilde{\alpha}$ with respect to minimal radius $r_0$:}}

We have analyzed that for $0<\sigma\leq50$ and $\sigma>50$ supposing $\frac{\omega_e}{\omega_\infty}=10^{-1}$, the deflection angle $\tilde{\alpha}$ increases as value of $r_0$ increases. A linear behavior of both the graphs has examined. We have also investigated that as $\sigma$ increases, angle decreases and vice versa.
We have reviewed that for $0<\sigma\leq50$ and $\sigma>50$ at $\frac{\omega_e}{\omega_\infty}=10^{-2}$, deflection angle $\tilde{\alpha}$ exhibited similar behavior as that for $\frac{\omega_e}{\omega_\infty}=10^{-1}$. We have also examined that $\tilde{\alpha}$ increases for larger $r_0$ and $\tilde{\alpha}$ decreases as $\sigma$ increases.  We have investigated a divergent behavior in both cases. We have evaluated that $\tilde{\alpha}$ has direct relationship with $r_0$ and inverse relation with $\sigma$.\\

Furthermore, we have also derived the deflection angle by using Keeton and Petters method which is the approximate form of GL and used spherically symmetric spacetime metric. For this purpose, we have found coefficients of PPN metric by comparing the expanded metric function with standard PPN metric. Later on, we have determined the coefficients of the bending angle and again by comparing them with general form of Schwarzschild metric to get final results given in Eq.(\ref{HF28}).

Lastly, we have calculated the value of deflection angle $\tilde{\alpha}$ for Kalb-Ramond traversable WH solution in the dark matter medium. Dark matter contained such particles which cannot absorb, emit or reflect light rays. For this purpose, we have calculated Gaussian optical curvature  Eq.(\ref{HF33}) and determined the deflection angle (\ref{HF36}).

In plasma, Keeton and Petters method and Dark matter mediums case, the results reduced to the non plasma case. The resulting equations depend upon the minimal radius $r_0$ and impact parameter $\sigma$ showing that deflection angle is directly proportional to $r_0$ and inversely to $\sigma$. The plots between angle and minimal radius or impact parameter demonstrates that when $r_0$ increases, bending angle also move towards infinity. Also we observe a convergent and divergent behavior w.r.t $\sigma$ and $r_0$ respectively. When we remove the effect of plasma and dark matter, the resultant expression becomes similar to that of Bumblebee traversable wormhole solution. In case of Keeton and Petters method only the first term becomes similar to the general case.

\acknowledgements
 A. {\"O}. and R. P. would like to acknowledge networking support by the COST Action CA18108 - Quantum gravity phenomenology in the multi-messenger approach (QG-MM).

\end{document}